# Thermal Stress Analysis of Liquid-Cooled 3D-ICs

Sakib Islam[1] and Ibrahim Abdel-Motaleb[2]
*Department of Electrical Engineering, Northern Illinois University , Dekalb, USA*
[1]aislam@niu.edu, [2]ibrahim@niu.edu

*Abstract*—It is known that 3D-ICs suffer from hot spot temperatures that can reach thousands of degrees, if they are not cooled to reasonable operating temperatures. The problem of hot spots is not limited to the high temperatures of the IC; thermal stress can also pose severe problems, even after cooling the chip. This study investigates thermal stress resulting from a 3D-IC hot spot with 20 W power dissipation. The IC is cooled using $SiO_2$ and diamond cooling blocks. The study is performed using three cooling liquids: water, Freon (R22), and Liquid Nitrogen (LN). As expected, the study shows that metal layers on the chip suffer from high thermal stress due to rising the chip temperature to values higher than the room temperature. It is also noticed that the stress becomes more severe, if cooling is done using LN. In fact, the stress exceeded the maximum tensile strength of aluminum, which means failure of the chip. This indicates that cooling 3D-IC may not ensure acceptable operation or reliability. Thermal stress must be investigated at both high and low temperatures to ensure high performance and acceptable reliability.

## I. Introduction

With the increasing density of ICs, two challenges have risen: the first is reaching the physical limit of the device size (reaching the end of Moore's law), and the second is the increase of the power density of the ICs to the point that the performance is adversely affected. To solve the first challenge, a new approach of integration where regular 2D-ICs are stacked on top of each other forming 3D-ICs [1]. The electrical power connections and signal transmission between the stacked ICs are done using metal vertical connections through holes, or vias. The 3D-ICs technology provides many advantages. First, the integration density can be increased by a factor equal to the number vertical stacks (for example, to 10,000%, if 100 2D-ICs are stacked on top of each other). Second, the total speed of the IC increases due to the short vertical connections between the stacks. Third, 3D-ICs allowed for integration of heterogeneous technology.

Although 3D-IC technology resulted in many advantages, it introduced serious problems. One of the problems is the introduction of power noise resulting from the need for huge power supplies and long signal delays [2]. The second, is the creation of hot spots, due to the high integration density. Hot spots temperatures may reach thousands of degrees that would destroy the IC in milliseconds. This is a very serious problem, especially for the next generation of IC where a power density reaching 1000 $kW/cm^2$ will be a requirement, according to ITRS [3].

To manage power dissipation from 3D ICs, several techniques have been proposed and implemented. Powerful fans, large heat sinks, Microelectromechanical Systems (MEMS)-based cooling technology, embedded micro channels [4,5], liquid immersion cooling [6], and microfluidic cooling using thermal TSVs [7] are among the cooling techniques utilized. Although the steady state temperature can be reduced to an acceptable operating temperature, there will be still time-dependent local temperature change [8]. Such dynamic change of the temperature can cause thermal stress and strain. The thermal energy produced from local hot spots can cause severe damage to the IC due to the thermal stress and strain. This mechanical stress is expected to affect the reliability and performance of the circuit, due to the likely chance of damage to the metal and insulator layers.

In this study, we investigated the thermal stress resulting from the dynamic temperature change of hot spots in liquid-cooled 3D ICs. The study was performed using advanced multi-physics numerical analysis program, Comsol$^{TM}$.

## II. Thermal Stress

The relationship between stress and strain is expressed by,

$$\boldsymbol{\sigma = E\varepsilon,} \qquad (1)$$

where $\boldsymbol{\sigma}$ is the uniaxial stress, $\boldsymbol{\varepsilon}$ is the strain or deformation, and $\boldsymbol{E}$ is Young's modulus [9]. Young's modulus, $\boldsymbol{E}$, gives the measure of stiffness of a solid material [10].

Thermal stress can be expressed by the following equation,

$$\boldsymbol{\sigma = E\alpha\Delta T = E\alpha(T_F - T_0),} \qquad (2)$$

where $\boldsymbol{\alpha}$ is the thermal expansion coefficient, $\boldsymbol{T_0}$ is initial temperature, $\boldsymbol{T_F}$ is final temperature, and $\boldsymbol{\Delta T = T_F - T_0}$ in kelvins (K). The temperature, $\boldsymbol{T_F}$ varies with respect to time due to subsequent heating and cooling and results in a temperature difference [11]. From equation (2), we can see that, for a specific material, the stress is proportional to the thermal expansion coefficient, young's modulus, and temperature difference. Since the materials used to build IC do not have the same thermal expansion, significant thermal stress may arise, adversely impacting the device performance and the circuit reliability [12].

## III. Design of The Cooling System

In this analysis, the test chip contains a heater to represent the hot spot in the 3D-ICs. The heat generated from a heater is assumed to be equal to the heat generated from the hot spot of the 3D ICs [8]. The test chip with the heater is shown in Fig. 1. In this chip, the substrate is a 200μm -thick silicon with an area of 5x5 $mm^2$; on top of the substrate, 0.30μm of $SiO_2$ is deposited; on top of the oxide layer, a 0.18 μm-thick serpentine-shaped tungsten layer with a total area of 1000 x 2000 μm$^2$ is deposited; another 0.15 μm layer of $SiO_2$ is deposited to cover the heater; on top of this oxide layer, two aluminum pads, with 125 μm x 250 μm area and 0.05 μm thickness, are deposited and connected to the two ends of the heater using two vias through the SiO2 layer. Finally, a 7000Å $SiO_2$ passivation layer is deposited.

Fig. 2 shows the cooling block structure which consists of blocks having an area of 40x40 $mm^2$. The inlet area is 2x38

mm². The thickness of the walls is 250 μm for the top and bottom and 1000 μm for the sides. The analysis was done using the two cooling blocks: one built using $SiO_2$ and the other using diamond. We used three different liquid coolants in the study: water, R22, and Liquid Nitrogen (LN).

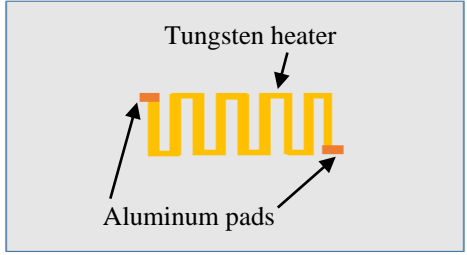

Fig.1: Serpentine Shaped Tungsten Heating Block on top of Si-Substrate.

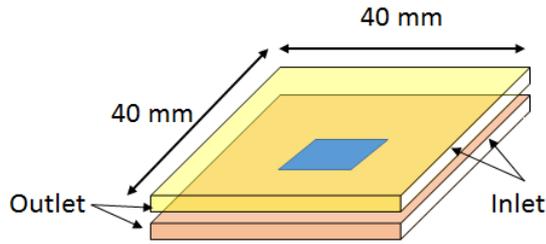

Fig. 2. Test chip sandwiched between two cooling blocks.

### IV. SIMULATION AND ANALYSIS

Comsol multiphysics program is used to conduct the thermal analysis. The emissivity values utilized in the simulation are 0.79 for $SiO_2$ and 0.63 for diamond [13,14]. The convective heat transfer coefficient is assumed to be 50 W/m²K for all the coolants and 10 W/m²K for air. The heater, representing a hot spot, is assumed to dissipate 20 W of heat.

Our analysis predicted that a 20W hot spot would elevate the chip temperature to 4050 K, if no cooling liquid is admitted, Fig. 3(a). This temperature would produce a thermal stress of 9418 MPa, as shown in Fig 3(b). This temperature is above the melting point of tungsten (3695 K), and the stress is beyond the maximum strength of the materials (980 MPa). This means that the IC will melt and evaporate before it can reach this temperature, and no structure will even remain to experience the stress.

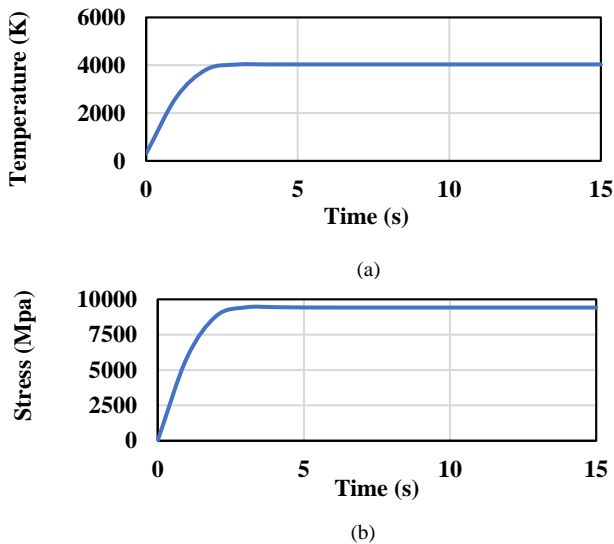

Fig.3: (a) Temperature & (b) Thermal Stress of the IC without Cooling.

To have a functioning IC, a cooling mechanism must be employed to cool the chip to the required operating temperature, within a reasonable time. Although, temperature can be reduced, an overshoot may take-place [8]. Overshoot of temperature may cause high thermal stress.

Thermal stress is investigated for a test chip sandwiched between two cooling blocks, as shown in Fig. 2. Temperature and the thermal stress are calculated for the first 15 s, after admitting the liquid coolants. With a 20W hot spot covering an area of 1000 x 2000 μm², the power density reaches 1000W/cm². Having this hot spot on a 25mm² chip results in a power dissipation of 80 W/cm² for the chip. In this analysis, the coolants were admitted at the same time the heater was powered up, where the system temperature was initially set to room temperature, 293K.

For $SiO_2$ cooling block, liquid was pumped at a velocity of 10 mm/s when the 20W heater is powered on, and temperature and thermal stress were calculated during the first 15 seconds. From Fig.4, we can observe that the temperature reaches an overshoot within 1 s, for water, R22, and liquid nitrogen cooling. The maximum temperature at the tungsten metal (hot spot) for water, R22, and liquid nitrogen reached 432K, 437K, and 375K and the corresponding coolants reached saturation at 432K, 416K, and 293K.

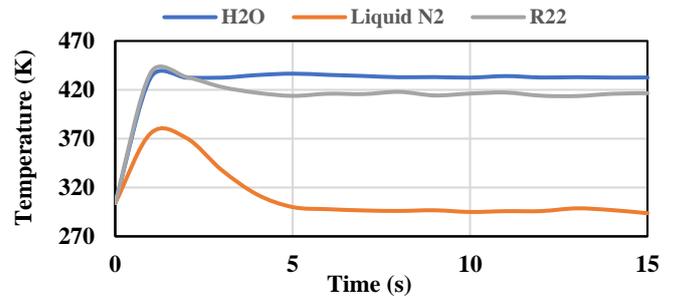

Fig. 4. Hot spot temperature using $SiO_2$ cooling blocks for water, Liquid $N_2$, and R22 coolant with 10mm/s inlet velocity.

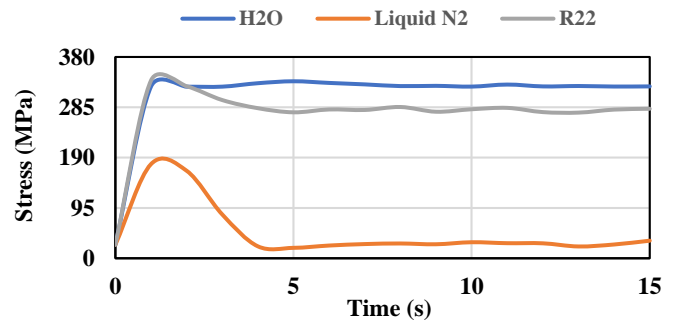

Fig. 5. Thermal Stress using $SiO_2$ cooling blocks on the IC for water, Liquid $N_2$, and R22 coolant with 10mm/s inlet velocity.

Fig.5 shows the corresponding thermal stress. Within 1s, the thermal stress rises from 0 MPa to maximum values of 324 MPa, 335 MPa, and 177 MPa for Water, R22, and Liquid $N_2$, respectively, then saturates to 324 MPa, 282 MPa, 33 MPa, as the temperature saturates. The figure shows that the stress value follows the temperature difference $|T_F-T_o|$ value ($T_0$=293K) as explained in Eq. (2). At 1s, $|T_F-T_o|$ has the values of 128K, 133K, 71K, for Water, R22, and Liquid $N_2$, respectively. This explains the reason for the high thermal stress for R22 compared to the other cooling cases at 1s. But as the temperature saturates, the difference from the original temperature goes down. This difference becomes smaller for R22 than water, leading to a lower stress. Cooling reduced the thermal stress to less than the ultimate tensile stress of tungsten of 980 MPa.

Aluminum 6061 is widely used in IC technology as interconnects and ohmic contacts. The ultimate tensile strength of that alloy is 310MPa [15]. For cooling using $SiO_2$ cooling block and velocity of 10mm/s, the thermal stress exerted on aluminum pads reaches up to 274 MPa for Water, 232 MPa for R22, and 68 MPa for LN. This high stress might lead to gradual breakdown of aluminum pads.

It appears that inlet velocity of 10 mm/s does not pump enough liquid to cool the IC down. Therefore, it is reasonable to assume that if the inlet velocity increased to 100mm/s, both the temperature and thermal stress will decrease. Fig. 6 shows that for 100 mm/s inlet velocity, the temperature saturates to 416K and 361K for water and R22, from initial temperature of 304 K, respectively. No significant overshoot of temperature for the two coolants take place. For LN, the temperature decreased to 207K within 1 s and saturates. This is due to the fact LN temperature is 77 K, and the velocity was high enough to pump enough LN to bring the temperature to a much lower value than the initial value much quicker.

Fig. 7 shows the thermal stress for an inlet velocity of 100 mm/s. The temperature difference, $|T_F - T_0|$ are 123 K, 68 K, and 86 K for water, R22, and LN, respectively. The stress can be due to tension or compression. The thermal stresses on the IC is related to $|T_F - T_0|$ and, accordingly, are found to be saturated at 303 MPa, and 161 MPa 233 MPa for water, R22 and LN, respectively. The study shows that thermal stress can exist for high as well as low final temperature. Hence. Deep cooling may case thermal degradation of the IC.

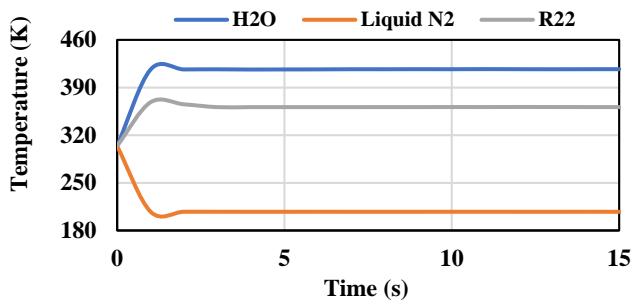

Fig. 6: Hot spot temperature after cooling using $SiO_2$ cooling blocks with 100 mm/s inlet velocity. Coolant used are water, R22, and LN.

Fig. 7 shows that the highest stress for tungsten results from water cooling. This is understood, since the temperature difference in this case is the highest. It is noticed that even though LN results in the lowest operating temperature, the stress is higher than the case with R22 cooling. This can be understood, if it is realized that the temperature difference, $|T_F - T_0|$ is higher in case of LN. Because the operating temperature for LN is lower than the initial value, the type of stress is the opposite (compression rather than tension).

Stress on aluminum is also investigated, for inlet velocity of 100 mm/s. The results show that the stress on aluminum pads are 241MPa, 107MPa, & 262MPa respectively for water, R22, and LN. Similar to the case of tungsten, R22 cooling provided the lowest stress, because the temperature difference is the lowest.

This results signified some issues. The first is that thermal stress exists not only because of higher operating temperatures, but also because of lower operating temperatures than the initial temperature. Therefore, if the intention is to operate at lower temperatures than room temperature, it may be advised to deposit metals at temperatures closer to the operating temperature. Second, the impact of metal elasticity and ultimate tensile stress are critical reliability factors that must be considered when designing 3D-ICs.

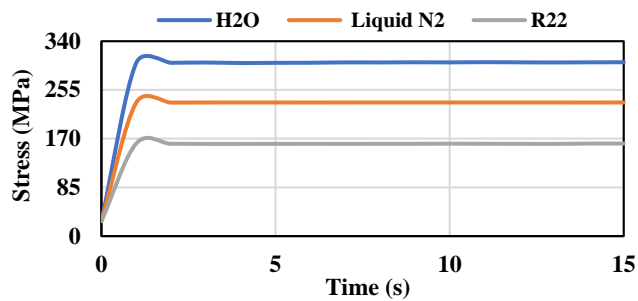

Fig. 7. Stress on tungsten when $SiO_2$ cooling blocks with 100 mm/s inlet velocity. Coolant used are water, R22, and LN.

Since $SiO_2$ is a poor thermal conductor, it is reasonable to assume that $SiO_2$ is responsible for the high temperature of the chip. For this reason, diamond, a highly thermal conductive material, is used instead to build the cooling block. Fig.8 shows the hot spot temperature for coolants pumped at 10 mm/s velocity. The saturation temperatures are found to be 325 K, 272 K, and 123 K, for water, R22, and LN, respectively. As can be seen from the figure, the saturation temperatures are much lower than when $SiO_2$ cooling blocks were used. The reduced temperature for water can be attributed to the higher thermal conductivity of diamond, which resulted in a faster dissipation of heat energy to the outside. For R22 and LN, the temperature is lower than the ambient temperature. This means, heat transfer should be from outside to inside, which in turn should raise the inside temperature for diamond blocks compared with $SiO_2$ block. But the opposite happens. The failure of heat to transfer from outside to inside is believed to be due to the creation of a cold domain outside the block. This domain acts as an isolating region that prevents heat from reaching the inside, with sufficient rate, to elevate the inside temperature.

The stress analysis for diamond blocks with 10 mm/s velocity is shown in Fig. 9. The thermal stress on tungsten was found to be 77 MPa for water, 58 MPa for R22, and 440 MPa for LN. As can be seen, the stress for water and R22 are low, since the temperature difference, $|T_F - T_0|$, are 21K and 17 K, respectively. For LN, the difference is 110K. This explains why the stress in case of LN is much higher. In all three cases, the stress is lower than the maximum tensile strength of tungsten. Similarly, for aluminum, the stress was found to be 42 MPa for water, 86 MPa for R22, and 443MPa for LN. Again the thermal stress for LN is high, but in this case it is higher than the maximum tensile strength of aluminum, which may cause metal fatigue.

When the inlet velocity increased to 100 mm/s, the temperature of the hot spot changed to 318 K for water, 260 K for R22, and 106 K for LN, as shown in Fig. 10. Fig. 11 shows the thermal stress which was found to be 60 MPa for water, 87 MPa for R22, and 482 MPa for LN. Similar to the case of 10 mm/s, LN results in high thermal stress, due to the large difference between the initial and final temperature values of the hotspot.

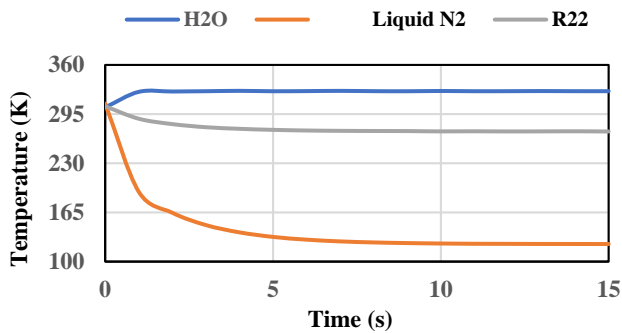

Fig. 8: Hot spot temperature after cooling using Diamond cooling blocks with 10 mm/s inlet velocity. Coolant used are water, R22, and LN.

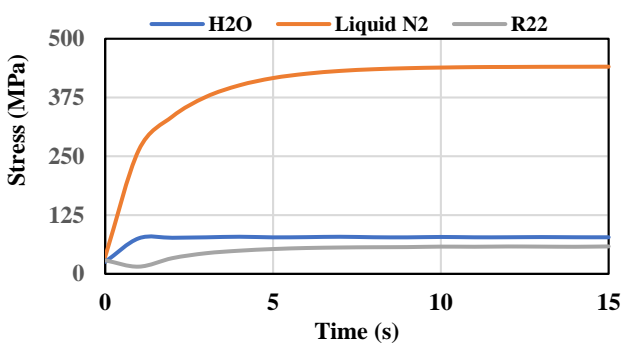

Fig. 9. Stress on the IC for using Diamond cooling blocks water, LN, and R22 coolant with 10 mm/s inlet velocity.

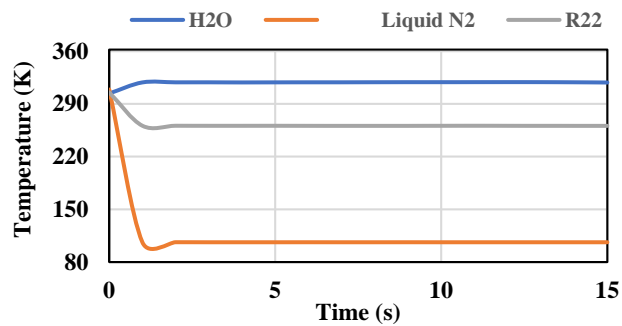

Fig. 10: Hot spot temperature for water, LN, and R22 coolant with 100 mm/s inlet velocity.

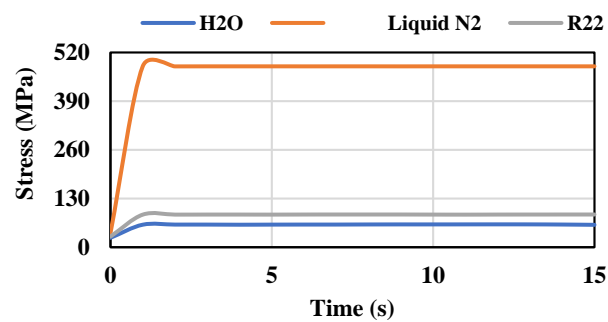

Fig. 11: Stress on the IC for water, LN, and R22 coolant with 100 mm/s inlet velocity.

The stress on the aluminum pads for 100 mm/s velocity is found to be 25 MPa for water, 118 MPa for R22, and 483 MPa for LN. Again, the stress due to LN cooling is the highest. The problem is that it is higher than the maximum tensile strength of the metal. This means the aluminum pads may break down and cause problem for the IC reliability, if cooled with LN.

## V. CONCLUSION

The stress analysis for next generation 3D-ICs have been conducted using diamond and $SiO_2$ cooling blocks. The result shows that, in order to ensure the stability of the IC thermal stress needs special consideration along with the operating temperature. The results also show that overcooling may cause metal breakdown. Therefore, the level of cooling should be considered, and the strength of materials used to build the IC should be evaluated. This study would not be complete without studying the material deformation due to the cooling process.